\documentclass[prb, twocolumn,superscriptaddress,showpacs]{revtex4}  
\usepackage{amssymb}
\usepackage{times}
\usepackage{amsmath}
\usepackage{bbm}
\usepackage{graphicx}
\usepackage{color}
\usepackage{enumerate}

\newcommand{\tmax}{\tau}

\begin{document}

\author{Nikolai Eurich}
\author{Martin Eckstein}
\author{Philipp Werner}
\affiliation{Theoretische Physik, ETH Zurich, 8093 Z{\"u}rich, Switzerland}

\title{Optimal ramp shapes for the fermionic Hubbard model in infinite dimensions}

\date{\today}

\hyphenation{}

\begin{abstract}
We use non-equilibrium dynamical mean field theory and a real-time diagrammatic impurity solver to study the heating associated with time-dependent changes of the interaction in a fermionic Hubbard model. Optimal ramp shapes $U(t)$ which minimize the excitation energy are determined for a noninteracting initial state and an infinitesimal change of the interaction strength. For ramp times of a few inverse hoppings, these optimal $U(t)$ are strongly oscillating with a frequency determined by the bandwidth. We show that the scaled versions of the optimized ramps yield substantially lower temperatures than linear ramps even for final interaction values comparable to the bandwidth. The relaxation of the system after the ramp and its dependence on the ramp shape are also addressed.
\end{abstract}


\maketitle

\section{Introduction}

Correlated electron systems play a central role in modern condensed matter physics.\cite{Imada98} 
While most theoretical studies have focussed on the equilibrium properties of these systems 
and many interesting phenomena remain to be understood, there is a growing effort to also 
address the non-equilibrium dynamics. These investigations are motivated by ongoing 
experimental progress in this field, in particular measurements of the relaxation dynamics 
in correlated materials by means of pump-probe spectroscopy,\cite{Perfetti2006a+Wall2009} 
and experiments on cold-atom systems.\cite{Bloch2008a} The latter provide an experimental 
realization of theoretical models such as the fermionic one-band Hubbard 
model,\cite{Joerdens2008a,Schneider2008a} and allow to explore the time-evolution after rapid 
changes in model parameters.\cite{Greiner2002b}

A computationally tractable method to simulate the time evolution of a correlated lattice model 
after a quench or external perturbation is nonequilibrium dynamical mean field 
theory.\cite{Schmidt2002, Freericks2006} This approach neglects spatial correlations, and local 
correlation functions and their dynamics are obtained from an appropriately defined quantum 
impurity model. The repeated numerical solution of this impurity model on 
the real-time (Keldysh) contour is the most challenging aspect of nonequilibrium 
DMFT calculations, but several perturbative\cite{EcksteinNCA} and exact\cite{Werner09noneq} 
impurity solvers have recently been developed to tackle this problem. 
Nonequilibrium dynamical mean field theory has been applied to 
interaction quenches in the Hubbard model\cite{Moeckel2008, Eckstein09quench,Eckstein10quench} 
as well as in the Falicov-Kimball model,\cite{Eckstein2008a} and to study the time evolution 
in the presence of external fields.\cite{Freericks2006,Freericks2008b,Tsuji08,Tsuji09,Eckstein10equench,Tsuji10} 
Interesting phenomena have been observed in these theoretical investigations, such 
as an apparent dynamical phase transition in the relaxation dynamics after an interaction 
quench.\cite{Eckstein09quench,Eckstein10quench}

A change in the interaction parameter affects the total energy and therefore the temperature 
of the system after equilibration, and nonadiabatic heating effects can be substantial.\cite{Zakrevski2009}  
The dynamical phase transition after a sudden interaction quench in the $T=0$, noninteracting Hubbard model,\cite{Eckstein09quench}  
for example, occurs at an energy which translates into a temperature of about 0.2 times the bandwidth, which is far higher than the temperatures 
for which a metal-insulator transition is found in equilibrium. While it is an interesting observation 
that an apparently sharp dynamical transition occurs in such a highly excited system, 
one would like to avoid the heating effect in other contexts.
In cold-atom systems, where much effort 
is devoted to realize a fermionic Hubbard model in the low-temperature regime, interactions 
and the optical lattice must be switched on in such a way that the heating is minimized. 
If there are constraints on the ramping time, it can become a subtle question whether 
an equilibrium state in a desired low-temperature phase can be reached at all by means of a ramp which starts from 
another, more easily preparable state of the system.\cite{Moeckel2010} And since in most cases one will be interested in preparing an equilibrium state of the interacting system, a second relevant question concerns the time-scale on which the system relaxes and its dependence on the ramp shape. 
Another example is the recently proposed conversion of repulsive interactions 
into attractive ones by means of external periodic electric fields (also essentially an interaction quench),\cite{Tsuji10} which raises the interesting 
possibility of AC-field induced superconductivity. 
Also in this context the heating and the thermalization time associated with the ramp protocol are important issues. 

The purpose of this theoretical study is 
to determine the optimal ramping procedure between two different 
parameter regimes of the Hubbard model, where ``optimal" means the passage in which the system is least 
excited.\cite{Barankov2008,Doria2010} 
In particular when the ramping time is restricted 
to only a few times the inverse hopping, one can expect a strong dependence of 
the excitation energy and of the relaxation dynamics on the ramping protocol. However, an unbiased optimization over 
the infinite space of ramp shapes is not possible, because even the 
computation of the excitation energy for a single ramp protocol requires 
considerable numerical effort. For ``small ramps'', 
on the other hand, 
in which the parameter is changed by only a small amount, one can resort to perturbation theory 
in the ramp amplitude to disentangle the influence of the ramp shape and many-body effects 
on the excitation energy.\cite{Eckstein10ramp} This allows us to perform an efficient 
optimization over a wider space of ramp shapes. In the present work we demonstrate that
ramp shapes which are optimized for such infinitesimal parameter changes yield a 
considerably lower excitation energy than the generic linear ramp or other simple 
ramping protocols, even when applied to ramps with an amplitude comparable to the bandwidth. We will furthermore demonstrate that a 
dynamical transition (associated with fast thermalization) also exists in the case of short ramps,
similar to what has been found for an interaction quench.\cite{Eckstein09quench, Eckstein10quench}
This observation suggests that a suitable choice of the ramp shape allows the preparation of thermal equilibrium states over a wide range of interaction values, within a switching time of only a few inverse hoppings.  

\begin{figure}[t]
\centering
\includegraphics[width=0.7\linewidth]{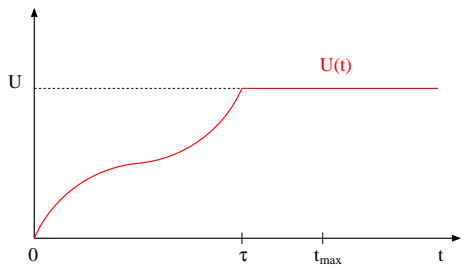}
\caption{Interaction ramp from $U(t=0)=0$ to $U(t=\tmax)=U$. The simulated time is $t_\text{max}>\tau$.}
\label{figramp}
\end{figure}

The specific model we consider is the one-band Hubbard model,
\begin{eqnarray}
H(t) = \sum_{ij\sigma}V_{ij}c_{i\sigma}^{\dag} c_{j\sigma} + U(t)\sum_i\Big(n_{i\uparrow} - \frac{1}{2}\Big)\Big(n_{i\downarrow} - \frac{1}{2}\Big),
\end{eqnarray}
with hopping amplitudes $V_{ij}$ and a time-dependent on-site repulsion $U(t)$. 
The hoppings $V_{ij}$ are chosen corresponding to a semi-elliptic density of states of bandwidth $4V$, 
$\rho(\epsilon)=\sqrt{4V^2-\epsilon^2}/(2\pi V^2)$ and we restrict our calculations to the paramagnetic phase 
of the half-filled model. We set $V=1$ as the unit of energy. Initially, the system is prepared in the 
noninteracting ground state ($U(t=0)=0$, $T(t=0)=0$). The interaction $U(t)$ is switched from $0$ to 
$U$ in a time $\tmax$ and then kept fixed at $U$ for $t>\tmax$ (see illustration in 
Fig.~\ref{figramp}). We explore different ramp shapes for switching on the interaction 
$U(t)$ and compute the resulting heating effect, which is quantified by the excitation energy.
Since the Hamiltonian is time-independent for $t>\tmax$, the excitation energy is constant for $t>\tmax$, 
even though other observables take time to equilibrate after the end of the ramp.

The rest of the paper is organized as follows. In Section~\ref{QMCsection} we briefly describe the real-time Monte Carlo impurity solver and the small adaptations needed to treat time-dependent interactions, in Section~\ref{perturbationsection} we recall the results of Ref.~\onlinecite{Eckstein10ramp} for the infinitesimal quench and show how $U(t)$ can be optimized based on these second order perturbation theory formulas. In Section~\ref{resultssection} we present nonequilibrium DMFT results for $U=1$ and $U=3$ and compare the optimized shapes to linear ramps. 
We will also discuss the time evolution of the momentum distribution function after the ramp and demonstrate that a fast thermalization occurs at a well-defined, but ramp shape dependent value of the interaction.
Section~\ref{outlooksection} is an outlook and conclusion.

\section{Weak-coupling Quantum Monte Carlo}
\label{QMCsection}
We compute the real-time evolution of the impurity model within the DMFT selfconsistent loop using a continuous-time Quantum Monte Carlo (CT-QMC) method \cite{Werner09noneq} which is based on an expansion of the partition function in powers of the interaction term.\cite{CTAUX} This method is free of systematic errors but restricted to relatively short times due to a dynamical sign problem. Nevertheless, it has proven useful for interaction quench calculations \cite{Eckstein09quench, Eckstein10quench, Tsuji10} in the Hubbard model and for the simulation of transport through  quantum dots.\cite{Werner10dot} 
Here, we briefly recapitulate the weak-coupling formalism to show that 
it can account for time-dependent interactions, simply by replacing weight factors associated with individual vertices by time-dependent factors. 
The remaining part of the DMFT self-consistency, i.e., the computation of lattice observables and 
lattice Green functions from the local Green function of the impurity system, is unchanged with respect 
to the interaction quench setup. A detailed description of these equations can be found in 
Ref.~\onlinecite{Eckstein10quench}.

We start by writing the partition function \(Z = Tr[e^{-\beta H}]\) as
\begin{eqnarray}
Z = \textit{Tr}\left[e^{-\beta H_1}T_{\textit{C}}e^{-\int_{\textit{C}}dtH_2}\right],
\end{eqnarray}
where $\int_{\textit{C}}$ denotes the integral along the Keldysh-contour $0\rightarrow t_\text{max}\rightarrow 0\rightarrow -i\beta$. In the simulations, we choose the length of the contour $t_\text{max}$ somewhat larger than the ramp time $\tau$.
The operators $H_1$ and $H_2$ correspond to the hopping and interaction part of the impurity Hamiltonian. Specifically, 
we use 
\begin{eqnarray}
H_2 &=& H_U - k(t)/t_\text{max},
\end{eqnarray}
with
\begin{eqnarray}
H_U = U(t)(n_{\uparrow}n_{\downarrow} - (n_{\uparrow} + n_{\downarrow})/2).
\end{eqnarray}
The nonzero real function $k(t)$ was introduced to enable an auxiliary field decomposition of the interaction term. Expansion of the contour-ordered exponential in powers of $H_U - k(t)/t_\text{max}$ and the application of the decoupling formula \cite{Rombouts99}
\begin{eqnarray}
1 - t_\text{max}H_U/k(t) &=& \frac{1}{2}\sum_{s=-1,1}e^{\gamma(t)s(n_{\uparrow} - n_{\downarrow})}, \\
\cosh(\gamma(t)) &=& 1 + \frac{t_\text{max}U(t)}{2k(t)}
\end{eqnarray}
at every interaction vertex results in an expression of the partition function as a sum over all possible collections of Ising spin configurations on the contour with weight
\begin{align}
&\omega(\{(t_1,s_1),(t_2,s_2),...,(t_n,s_n)\}) =\nonumber\\ 
&\hspace{3mm}(-i^{n_-})(i^{n_+})(k(t)dt/(2t_\text{max}))^{n_- + n_+}\prod_{\sigma}\det N_{\sigma}^{-1}.\hspace{3mm}
\end{align}
Here, $n_+$ and $n_-$ are the number of spins on the forward and backward branch of the contour, and we have used the fact that in a noninteracting initial state there are no spins on the imaginary-time branch. The matrices \(N_{\sigma}^{-1}\) are determined by the location of the interaction vertices on the contour \textit{C} and are given by
\begin{eqnarray}
N_{\sigma}^{-1} = e^{S_{\sigma}} - (iG_{0,\sigma})(e^{S_{\sigma}}-I),
\end{eqnarray}
with \(G_{0,\sigma}\) the bath Green's function and \(e^{S_{\sigma}} = \textrm{diag}(e^{\gamma(t_1)s_1\sigma},...,e^{\gamma(t_n)s_n\sigma})\).
In the actual calculations, we choose $\gamma(t)=\gamma$ time-independent,
so that the time dependence of the interaction manifests itself only in a time dependence of the parameter $k(t)=\frac{1}{2} t_\text{max}U(t)/(\cosh(\gamma)-1)$. 

The sampling procedure consists of generating all spin configurations on the contour \textit{C} through random insertions and removals of spins. During the sampling, we measure the quantity
\begin{eqnarray}
&&X_{\sigma}(s_1,s_2) =\nonumber\\
&&\,\,\,\Big\langle i\sum_{i,j=1}^n \delta_C(s_1,t_i)[(e^{S_{\sigma}}-1)N_{\sigma}]_{i,j}\delta_C(s_2,t_j)\Big\rangle_{MC}.
\end{eqnarray}
The impurity Green's function is then obtained as 
\begin{align}
&G_{\sigma}(t_1,t_2) = G_{0,\sigma}(t_1,t_2) \nonumber\\
&+ \int_{\textit{C}}ds_1\int_{\textit{C}}ds_2G_{0,\sigma}(t_1,s_1)X_{\sigma}(s_1,s_2)G_{0,\sigma}(s_2,t_2).
\end{align}

\section{Pertubative analysis for small ramps}
\label{perturbationsection}

\subsection{General considerations}

In this section we briefly recall the perturbative analysis of Ref.~\onlinecite{Eckstein10ramp} 
for an infinitesimal interaction ramp. To be specific, we consider a ramp of the form 
\begin{equation}
U(t) = U \,r(t/\tmax),
\end{equation}
where $r(x)$ is the ramp-shape function which satisfies $r(x)=0$ for $x\le 0$ and $r(x)=1$ for
$x\ge 1$. The excitation energy per lattice site is defined as 
\begin{eqnarray}
\Delta E(\tmax) = E(\tmax) - E_0(\tmax),
\end{eqnarray}
where $E_0(\tmax)$ is the groundstate energy per lattice site 
of the model with interaction $U$,
and $E(t) = \langle H(t) \rangle $ is the energy expectation value of the system
with time-dependent interaction.
After expanding in powers of $U$, the second-order result for the excitation energy can be decomposed
into contributions from a ramp spectrum $F(x)$, which depends only on the ramp shape 
but not on the properties of the system, and from an excitation density $R(\omega)$, which 
depends on the system but not on the ramp shape,\cite{Eckstein10ramp}
\begin{eqnarray}
\Delta E(\tmax) &=& U^2 \mathcal{E}(\tmax) + O(U^3), \\
\label{integral-E}
\mathcal{E}(\tmax) &=& \int_0^{\infty} \frac{d\omega}{\omega}R(\omega)F(\omega \tmax), \label{Eex}\\
F(x) &=& \left | \int_0^1\!ds\,r^{\prime}(s)e^{ixs}\right |^2.\label{defF}
\end{eqnarray}
In this expression, the excitation density is defined by the Lehmann representation, 
\begin{equation}
R(\omega)=\frac{1}{L}\sum_{n\neq 0} |\langle \phi_n |W| \phi_0\rangle |^2 \delta(\omega-E_n+E_0),
\end{equation}
where $W=(n_\uparrow-\tfrac12)(n_\downarrow-\tfrac12)$ is the operator which couples to 
the time-dependent parameter $U(t)$ in the Hamiltonian, $L$ is number of lattice sites, 
and $|\phi_n\rangle$ and $E_n$ are the eigenfunctions and eigenvalues of the Hamiltonian. 
 The function $R(\omega)$ can be evaluated
easily for the Hubbard model with $U=0$, leading to 
\cite{Eckstein10ramp}
\begin{align}
R(\omega) = &
\int_{-\omega}^0
\!\!\!\!d\epsilon\,\rho(\epsilon)
\int_0^{\epsilon+\omega}
\!\!\!\!d\mu\,\rho(\mu-\omega-\epsilon)
\int_0^{\mu}
\!\!\!\!d\nu\rho(\nu)\rho(\mu-\nu). \label{defR}
\end{align}

\begin{figure}[t]
\centering
\includegraphics[width=1.0\linewidth]{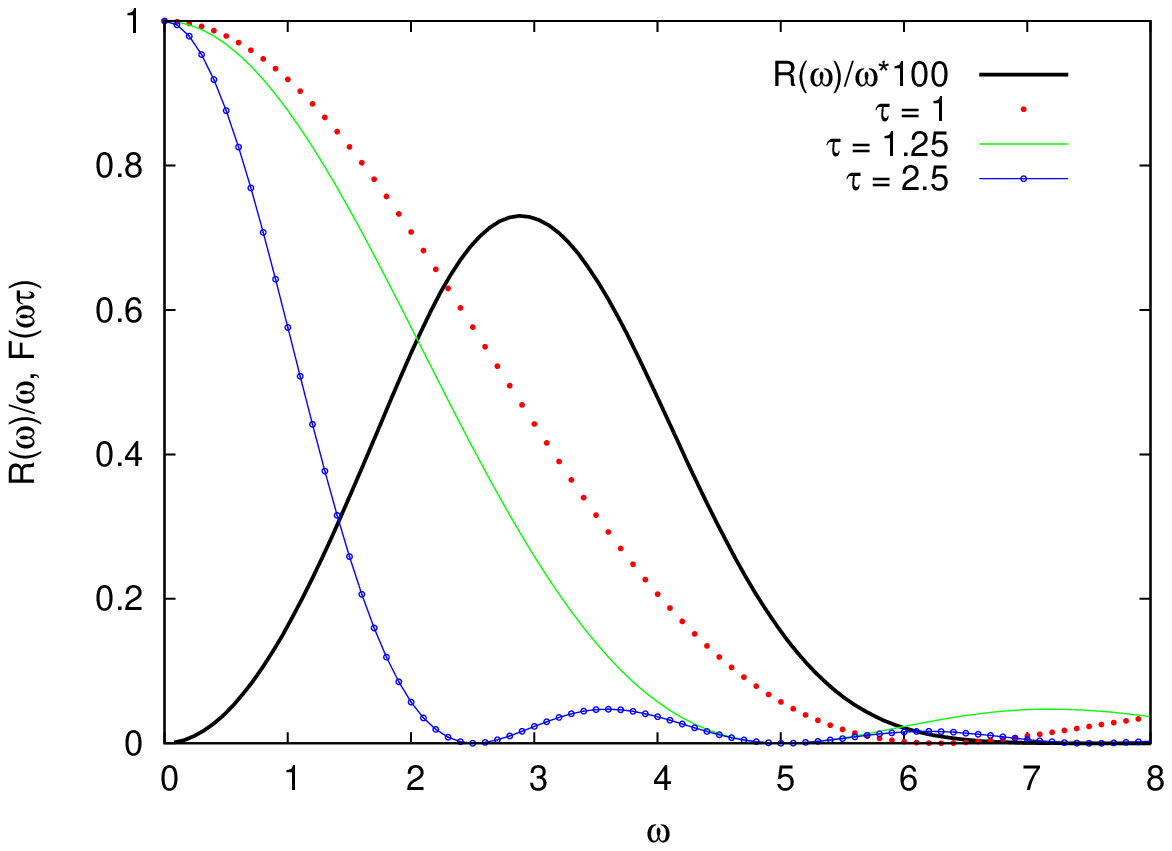}
\includegraphics[width=1.0\linewidth]{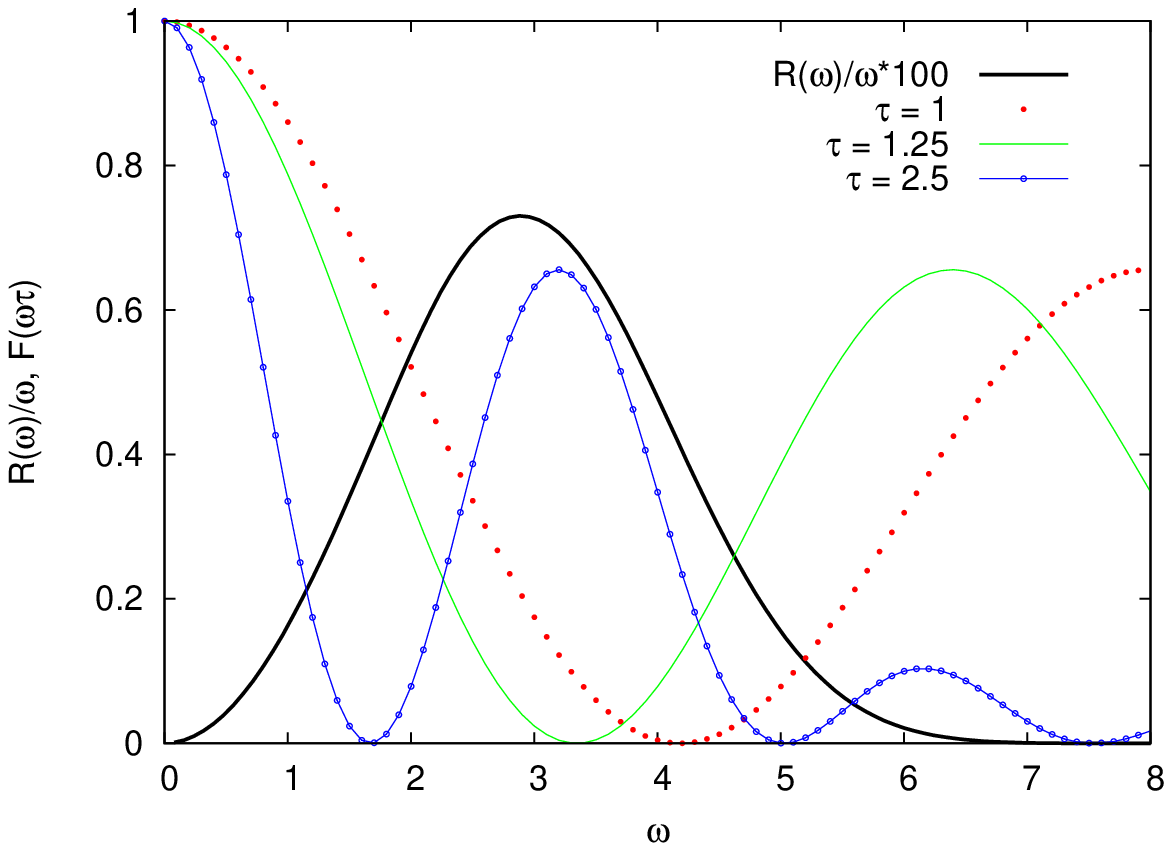}
\caption{Spectral density $R(\omega)/\omega$ and ramp spectra $F(\omega \tmax)$ for linear ramps with $\tmax= 1$, 1.25, 2.5 (top panel), as well as ramps of the form $r(x)= x+0.2\sin(2\pi x)$ and $\tmax= 1$, 1.25, 2.5 (bottom panel).
}
\label{spectraldensity}
\end{figure}

In Fig.~\ref{spectraldensity} we plot the function $R(\omega)/\omega$ for the semi-elliptical 
density of states together with ramp spectra for different ramp shapes. At $U=0$, the only 
relevant energy scale is the hopping $V$ $(=1)$: $R(\omega)/\omega$ has a peak near $\omega\approx 3$ 
and vanishes for $\omega>8=2\cdot\text{bandwidth}$. According to Eq.~(\ref{Eex}), a small overlap of the excitation  
density and ramp spectrum results in a small excitation energy. If the ramp spectrum $F(x)$
falls off rapidly at large $x$, the main weight of $F(\omega\tmax)$ and the main contribution
to the integral in Eq.~(\ref{integral-E}) comes from  frequencies $\omega \lesssim 1/\tmax$.
Using the asymptotic form 
$R(\omega) \propto \omega^3$ for small $\omega$ one can prove that the system approaches 
the adiabatic limit $\tmax\to\infty$ with a power law behavior $\Delta E(\tmax) \propto 1/\tmax^3$, provided 
that the high-frequency tail of $F(x)$ falls off faster than $1/x^3$. For the linear ramp $r(x)=x$, for 
which $F(x)$ falls off as $1/x^2$, one finds $\Delta E(\tmax) \propto 1/\tmax^2$.\cite{Eckstein10ramp}

In the present paper we are interested in the excitation energies for ramp times which are
so short that the asymptotic power law behavior $\Delta E(\tmax) \propto 1/\tmax^\eta$ 
does not yet hold, and we attempt to minimize $\Delta E(\tmax)$ in Eq.~(\ref{integral-E}) 
with respect to the ramp shape $r(x)$. The reduction of the excitation energy can be 
understood as a consequence of a suppression of the ramp spectrum $F(\omega\tau)$ in the 
frequency range around the maximum of $R(\omega)/\omega$, which is achieved by suitably 
designing the ramp $r(x)$. Clearly, the optimal ramp shape 
will strongly depend on the ramp time $\tmax$. For example, for the ramp 
$r(x)= x+0.2\sin(2\pi x)$ the function $F(\omega\tmax)$ is suppressed around 
$\omega=3$ for $\tmax\approx 1.25$, while for $\tmax=2.5$ the same ramp shape is apparently not favorable (Fig.~\ref{spectraldensity}b).

\subsection{Minimization in a tent basis}

To find the optimal ramp-protocol $r(x)$ which minimizes the excitation energy $\mathcal{E}(\tmax)$ 
given some constraints on $r(x)$, we discretize the interval $[0,1]$ into an equidistant mesh of 
$N+2$ points $x_k= k\Delta x$ ($k=0, \ldots, N+1$ ; $\Delta x=1/(N+1)$), and consider ramp functions which 
lineraly interpolate between the values $r(x_k)= x_k+c_k$. The numbers $c_k$ measure the deviations 
from the linear ramp $r(x)=x$, so $c_0=c_{N+1}=0$. Equivalently, this means that the function $r(x)$ is 
expanded in a basis of tent-shaped functions $\phi_k(x)$,
\begin{eqnarray}
r(x) = x + \sum_{k=1}^Nc_k\phi_k(x),
\end{eqnarray}
with
\begin{eqnarray}
\phi_k(x) = 
  \begin{cases} 
   \frac{x - x_{k-1}}{x_k - x_{k-1}} & \text{if } x \in [x_{k-1},x_k], \\
   \frac{x_{k+1}-x}{x_{k+1}-x_k} & \text{if } x \in [x_{k},x_{k+1}], \\
   0 & \text{otherwise }.
  \end{cases}
\end{eqnarray}
The first derivavtive of the ramp function \(r(x)\) is given by
\begin{eqnarray}
r'(x) = 1 + \sum_{k=1}^Nc_k\phi_k'(x)
\end{eqnarray}
with
\begin{eqnarray}
\phi_k'(x) = 
  \begin{cases} 
   \frac{1}{\Delta x} & \text{if } x \in [x_{k-1},x_k], \\
   \frac{-1}{\Delta x} & \text{if } x \in [x_{k},x_{k+1}], \\
   0 & \text{otherwise}.
  \end{cases}
\end{eqnarray}
The calculation of $F(\omega \tmax)$ requires the evaluation of the integral 
\begin{align}
&\int_0^1r'(s)e^{i\omega \tmax s}ds 
= \frac{1}{i\omega \tmax}\left(e^{i\omega \tmax} - 1\right) \nonumber\\
&\hspace{15mm}+ \frac{2(1 - \cos(\omega \tmax\Delta x))}{i\omega \tmax\Delta x}\sum_{k=1}^Nc_ke^{i\omega \tmax x_k}.\label{intrprime}
\end{align}
Multiplying the right hand side of Eq.~(\ref{intrprime}) with its complex conjugate and remembering that $F(x)$ is real gives 
\begin{align}
\mathcal{E}(\tau) &= \int_0^{\infty}d\omega\frac{R(\omega)}{\omega} F(\omega \tmax) \nonumber\\
&= \text{const} + \mathbf{c}^T\mathbf{f} + \mathbf{c}^T\mathbf{M}\mathbf{c},
\end{align}
where the coefficients $c_k$ are written as an $N$-component vector $\mathbf{c}$,
and the $N$-component vector $\mathbf{f}$ and the $N\times N$ matrix $\mathbf{M}$ are 
given by
\begin{align}
f_k
&=
\int_0^{\infty}d\omega\frac{R(\omega)}{\omega}\frac{4(1 - \cos(\omega \tmax\Delta x))} {\left | \omega \tmax\right |^2\Delta x} \nonumber\\
&\hspace{5mm}\times\left[(\cos(\omega \tmax (1- x_k)) - \cos(\omega \tmax x_k)\right], \label{felements}\\
M_{kk'} 
&=
\int_0^{\infty}d\omega\frac{R(\omega)}{\omega}\frac{4(1 - \cos(\omega \tmax \Delta x))^2}{\left | \omega \tmax \right |^2\Delta x^2}\nonumber\\
&\hspace{5mm}\times\cos(\omega \tmax(x_k - x_{k'})).\label{elements}
\end{align}
The minimization of this quadratic problem can be performed using standard techniques.

\subsection{Unconstrained optimization and symmetries}

In order to get a qualitative understanding of the optimal ramp shapes we first note the following symmetry properties of \(\mathbf{M}\) and \(\mathbf{f}\): 
\begin{enumerate}
\item $\mathbf{M}$ is symmetric, i.e. $M_{i,j}  = M_{j,i}$,
\item $M_{i,j}$ is constant on each subdiagonal $|i-j|=k$,
\item $\mathbf{f}$ is antisymmetric, i.e. $f_i = -f_{N +1 - i }$.
\end{enumerate}

The optimal, unconstrained ramp shape can be found by solving $\partial \mathcal{E}(\tau; \{c_i\})/\partial {c_i} = 0$ and therefore, using the symmetry of $\mathbf{M}$
\begin{eqnarray}
2\mathbf{M}\mathbf{c} + \mathbf{f}  = 0. \label{unconstrained}
\end{eqnarray}
It follows that the solution $\mathbf{c}$ is {\it antisymmetric}, i.e.  $c_i = -c_{N +1 - i }$.
In fact, setting $N = 2k$, the first $i = 1, \ldots, k$ components of Eq.~(\ref{unconstrained}) are
\begin{eqnarray}
-f_i = \sum_{j=1}^k \left( M_{i,j}c_j + M_{i,2k+1-j}c_{2k+1-j} \right).
\end{eqnarray}
Using the symmetry $f_i = -f_{2k+1-i}$ we get
\begin{align}
0 &= \sum_{j=1}^k\left(M_{i,j} + M_{2k+1-i,j}\right)c_j \nonumber\\
&\hspace{10mm}+ \left(M_{i,2k+1-j} + M_{2k+1-i,2k+1-j}\right)c_{2k+1-j}.
\end{align}
The symmetry properties of $\mathbf{M}$ thus imply
\begin{eqnarray}
0 = \sum_{j=1}^k\left(M_{i,j} + M_{i,2k+1-j}\right)(c_j + c_{2k+1-j}),
\end{eqnarray}
which can only be satisfied for all $i=1, \dots, N$ if $c_j = -c_{2k-j+1}$. 

\section{Results}
\label{resultssection}
\subsection{Small ramp amplitude}

\begin{figure}[t]
\centering
\includegraphics[width=0.95\linewidth]{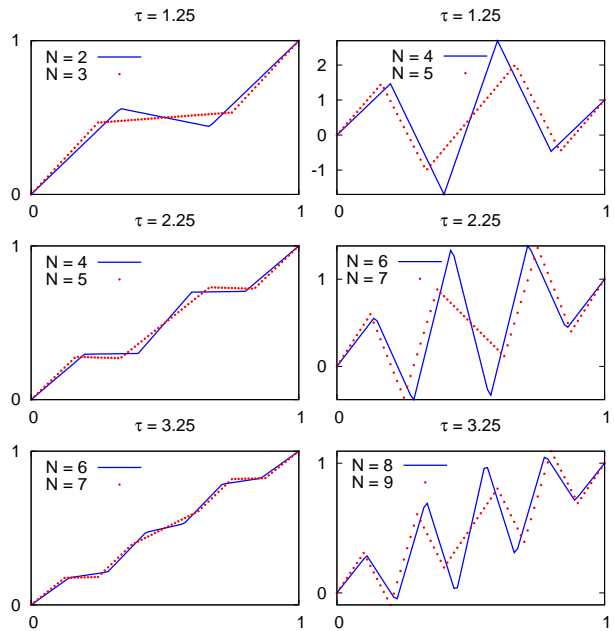}
\caption{Optimized unconstrained ramp shapes for $\tmax = 1.25$, $2.25$, $3.25$ and indicated number of basis functions ($N$).}
\label{figunconstrained}
\end{figure}

We first consider unconstrained paths and solve Eq.~(\ref{unconstrained}) for different values of $\tmax$. As illustrated in Fig.~\ref{figunconstrained}, the optimal paths oscillate around the linear ramp, and the number of oscillations
increases with increasing $\tmax$. Thus, the somewhat unexpected result is that paths which minimize the excitation energy, at least according to the perturbative formula (\ref{Eex}), may involve excursions to positive and negative interaction values which are much larger in absolute value than the final interaction $U$.

Note that the uncontrained optimization as decribed here becomes numerically unstable for large $N$. 
This is because one can always add to the ramp a highly oscillating component $\delta r(x)$ whose ramp spectrum $\delta F(x)$ lies almost completely outside the support of the excitation density  
$R(\omega)$, and thus does not influence the excitation energy. In other words, there are directions 
in the parameter space along which the excitation energy hardly changes. The corresponding eigenvalues of the 
matrix $\mathbf{M}$ are almost zero, and the linear equation (\ref{unconstrained}) becomes ill
conditioned for large $N$. 

\begin{figure}[t]
\centering
\includegraphics[width=1\linewidth]{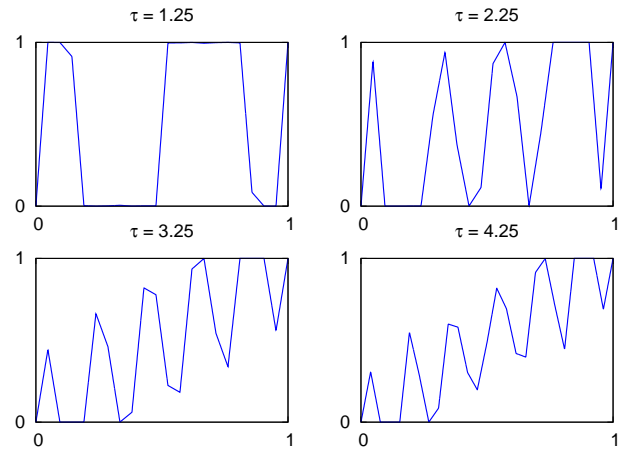}
\caption{Optimized constrained ramp shapes for $\tmax = 1.25$, $2.25$, $3.25$, $4.25$ and $N=20$.}
\label{figconstrained}
\end{figure}

\begin{figure}[t]
\centering
\includegraphics[width=1.0\linewidth]{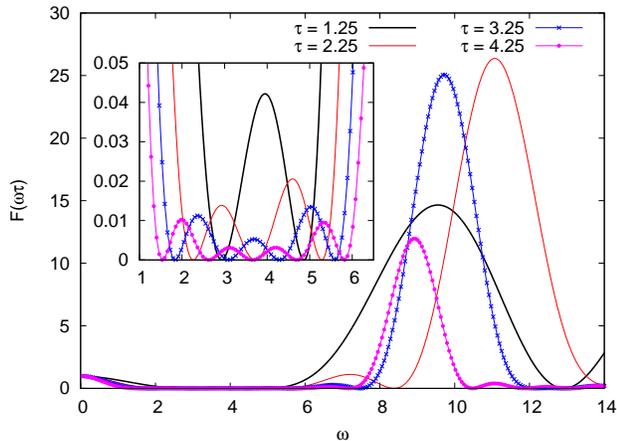}
\caption{Ramp spectra $F(\omega\tau)$ for the optimized constrained ramp shapes in Fig.~\ref{figconstrained} ($\tmax = 1.25$, $2.25$, $3.25$, $4.25$ and $N=20$). The inset shows a close-up view of the energy range $1\le \omega \le 6.5$ in which the excitation density $R(\omega)$ is large and hence  the ramp spectra are strongly suppressed.}
\label{F_constrained_ramps}
\end{figure}

However, optimal ramps with large amplitude oscillations and sign changes 
may be difficult to realize in experiments. 
We therefore also compute optimal ramps with the constraint $0\le r(x) 
\le 1$, using the reflective Newton method \cite{Coleman96} implemented in MATLAB.\cite{Matlab} The resulting paths for different $\tmax$ are 
shown in Fig.~\ref{figconstrained}.
It is evident from these results that the optimal ramp shapes are characterized by a roughly constant  
oscillation frequency $\omega_0=2\pi n_\text{osc}/\tau \approx 9-10$. To gain some insight into the origin of these oscillations we plot in Fig.~\ref{F_constrained_ramps} the ramp spectra $F(\omega\tau)$ for the ramps shown in Fig.~\ref{figconstrained}.
The inset shows a close-up view of the optimal ramp spectra in the frequency range where $R(\omega)$ is large. We see that $F(\omega\tau)$ is optimized in such a way that the overlap with $R(\omega)$ is minimal. The main panel shows the ramp spectra in the range $0\le\omega\le14$. A large peak in $F(\omega\tau)$ is evident at $\omega\approx 9.5$ ($\tau=1.25$), 11 ($\tau=2.25$), 9.75 ($\tau=3.25$) and 9 ($\tau=4.25$),  
i.e., just above the largest value of $\omega$ for which $R(\omega)>0$. We can understand from Eq.~(\ref{defF}) that a sharp peak in $F(\omega\tau)$ at $\omega\approx \omega_0$ corresponds to  an oscillating $r'(s)\sim \cos(\omega_0\tau s)$. So, the frequency which appears in the optimal ramp shapes is determined by the support of the function $R(\omega)$, which itself is defined by the density of states $\rho(\omega)$. In the case of a symmetric $\rho(\omega)$ considered here, the support of $R(\omega)$ is twice the bandwidth (Eq.~(\ref{defR})). The optimal ramp shapes have therefore an oscillating component with an oscillation frequency roughly given by $\omega_0\approx 2\cdot\text{bandwidth}$.

\subsection{Monte Carlo results for larger $U$}

In this subsection, we use Monte Carlo simulations to compute the heating effect for ramps with amplitudes beyond the perturbative regime, and compare the heating produced by linear ramps to the heating produced by ramps which were optimized for infinitesimal ramp amplitudes. 

\begin{figure}[t]
\centering
\includegraphics[width=1.0\linewidth]{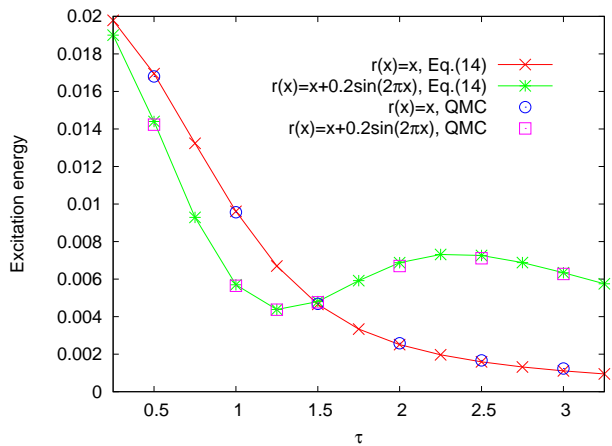}
\caption{Comparison of the DMFT data (QMC) to the results obtained from the perturbative analysis (Eq.~\ref{Eex}) for both linear and sinusoidal ramps to $U=1$ and for various ramp times $\tmax$. The interacting ground state was estimated to be $E_0=0.6195$.}
\label{figsmallu}
\end{figure}

First, we would like to confirm the validity of the perturbative analysis for ramps 
to small interaction. In Fig.~\ref{figsmallu} we plot the excitation energy computed by means of nonequilibrium DMFT for different ramps to $U=1$. The red line with crosses, and the green line with stars show the result of Eq.~(\ref{Eex}) for linear ramps and ramps of the form $r(x)=x+0.2\sin(2\pi x)$, respectively. Blue circles and pink squares show the excitation energy obtained from the DMFT calculation for an interacting ground state energy $E_0=0.6195$. This value of $E_0$ gives the best agreement between analytical and DMFT results, and is consistent within error bars with the ground state energy $E_0^\text{DMFT}(U=1)=0.618(2)$ estimated from equilibrium DMFT calculations. Hence, for ramps to $U=1$ the formula (\ref{Eex}) gives accurate excitation energies. The ramp spectra $F(\omega \tmax)$ for several values of $\tau$ are plotted in Fig.~\ref{spectraldensity} and explain the nonmonotonic behavior of the excitation energy in the case of  $r(x)=x+0.2\sin(2\pi x)$.

\begin{figure}[t]
\centering
\includegraphics[width=1.0\linewidth]{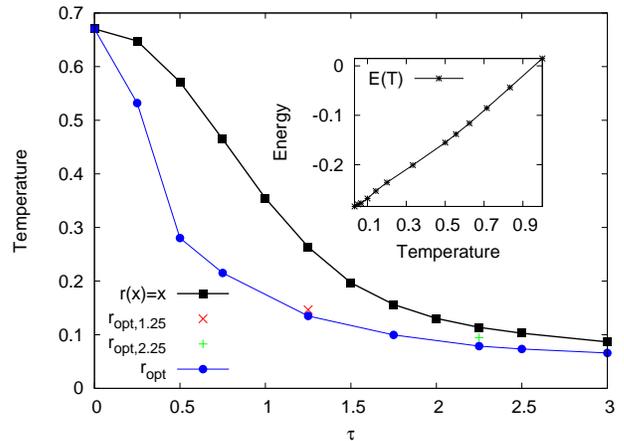}
\caption{Effective temperature for linear ramps (black squares) and optimized constrained ramps (blue circles) to $U = 3$ as a function of ramp-up time $\tmax$. The inset shows the relationship between the energy after the ramp and the temperature of an equilibrium model with interaction $U$ and the same total energy.}
\label{figlinramp}
\end{figure}

A nontrivial question is whether the ramp shapes optimized using Eq.~(\ref{Eex}) yield low excitation energies also for larger values of $U$. In the following we will consider ramps to $U=3$. This value is close to the critical interaction strength $U_c\approx 3.2$ for instantaneous quenches, where the system was found to thermalize within a time of less than two inverse hoppings.\cite{Eckstein09quench, Eckstein10quench} 
To quantify the heating effect we 
translate the excitation energy into an effective temperature (inset of Fig.~\ref{figlinramp}), which is defined as the temperature for which an equilibrium system with interaction $U$ has total energy $E(\tmax)$. 

\begin{figure}[t]
\centering
\includegraphics[width=1\linewidth]{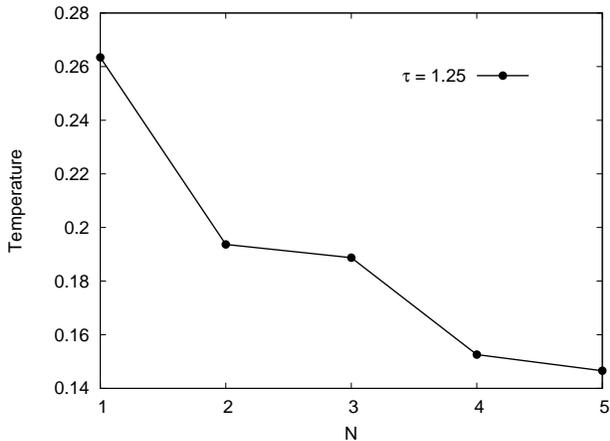}
\caption{Effective temperatures 
for optimized unconstrained ramps (see top panels of Fig.~\ref{figunconstrained}) with $N$ basis functions. $U=3$, $\tmax=1.25$.}
\label{figteffunconstrained}
\end{figure}

The effective temperature for linear ramps to $U=3$ is plotted as a function of ramp time in the main panel of Fig.~\ref{figlinramp} (black line with squares).  We see that the temperature after the quench drops rapidly with $\tmax$ until about $\tmax\approx 1.5$ and then more slowly for longer ramp times.
The times $\tmax$ which are accessible with Monte Carlo are not 
sufficient to determine the asymptotic behavior of the excitation energy ($\tmax\to\infty$), but it is expected to be 
$\Delta E(\tmax) \sim 1/\tmax^2$ based on the perturbative anaysis.
Figure~\ref{figteffunconstrained} shows the effective temperatures for $\tmax=1.25$ 
obtained from the optimized unconstrained ramps for different numbers $N$ of basis functions (see top panels of Fig.~\ref{figunconstrained}). The lowest temperature for $\tmax=1.25$, $N=5$ and $\tmax=2.25$, $N=7$, are indicated by the red and green crosses in Fig.~\ref{figlinramp}. For short ramp times ($\tmax=1.25$), the effective temperature can be reduced by about a factor of $2$ (compared to linear ramps) if an oscillating ramp shape is used. For $\tmax=2.25$, the reduction is only about 10\%.
The constrained optimized ramps yield comparable reductions in the  effective temperature. In Fig.~\ref{figlinramp} the blue line with circles shows the effective temperature for $N=25$ basis functions and the constraint $0\le r(x)\le 1$. 

\begin{figure}[t]
\centering
\includegraphics[width=0.95\linewidth]{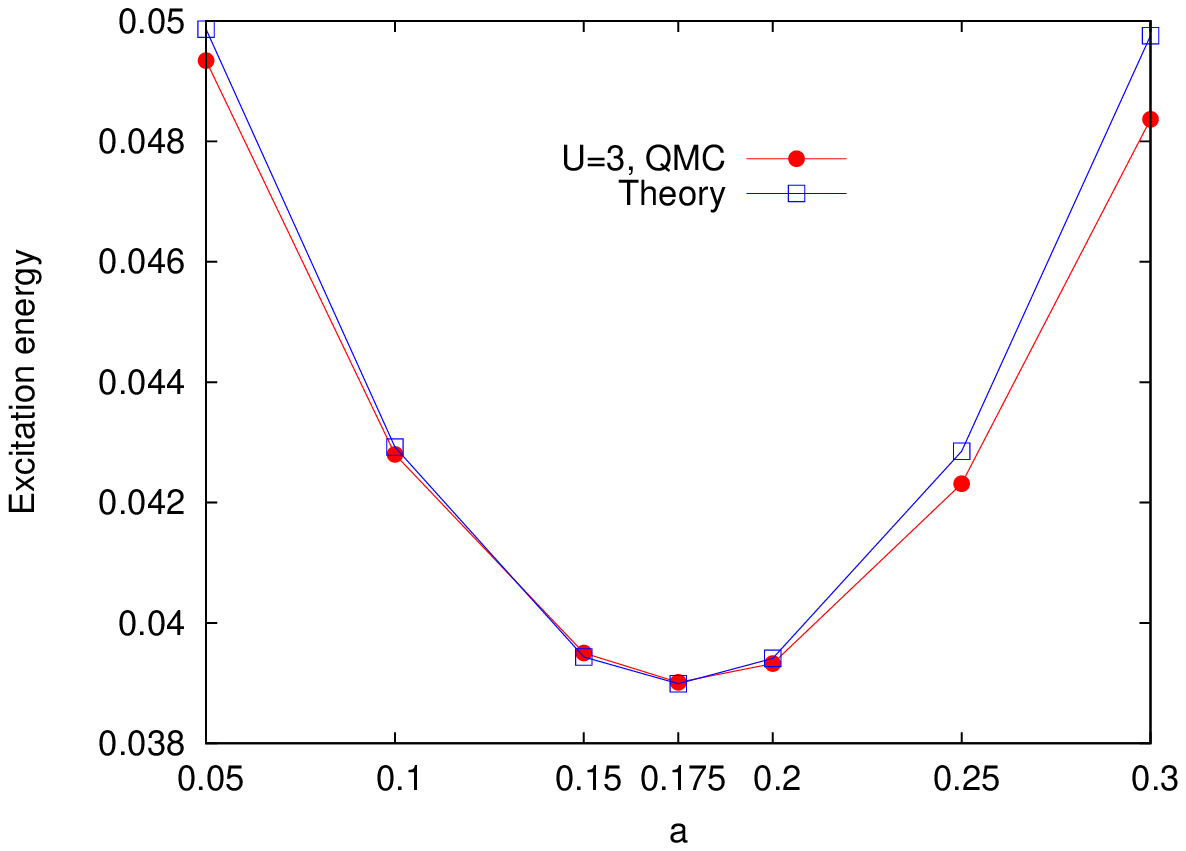}
\includegraphics[width=0.95\linewidth]{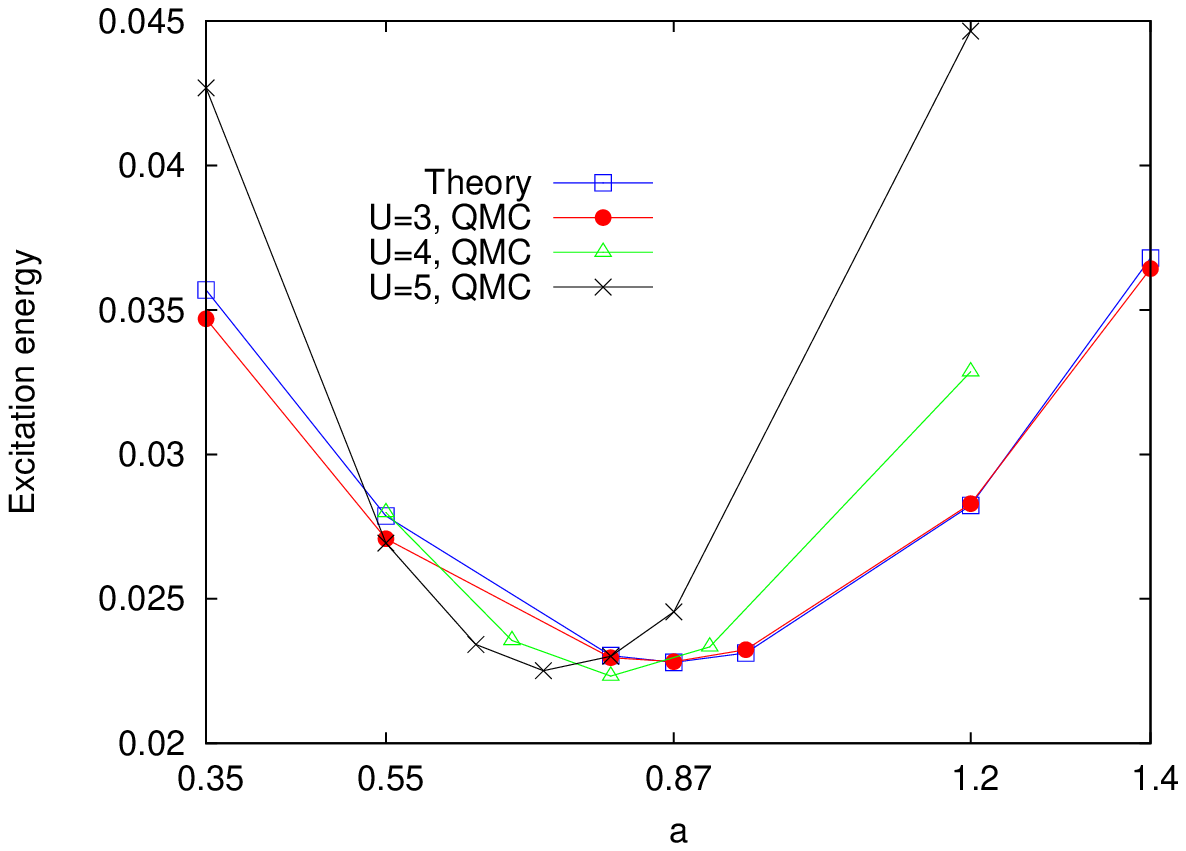}
\caption{Excitation energies for the one-parameter model $r(x)=x+a\sin(2\pi x)$ (top panel), and $r(x)=x+a\sin(4\pi x)$ (bottom panel) for $\tmax = 1.25$, $U=3$.
The blue line with squares shows the result from second order perturbation theory, while the red line with circles shows the Monte Carlo data shifted in such a way that the minima of the curves coincide. In the lower panel we also plots Monte Carlo results (with arbitrary energy offset) for $U=4$ and $U=5$.}
\label{figoneparameter}
\end{figure}

\subsection{Accuracy of the perturbative results}
 
The results shown in Fig.~\ref{figlinramp} demonstrate that the perturbative analysis of Section~\ref{perturbationsection} may be used to compute  ramp shapes with considerably lower excitation energy than linear ramps.
On the other hand, it is not yet clear how quantitatively accurate the estimated energies are in the case of ramps to intermediate or strong $U$. If the difference between the true and the estimated excitation energy is large and the error depends sensitively on the ramp shape, then the true optimal ramp might look very different from the shape obtained by our procedure. To test the reliability of our estimated excitation energy and its dependence on the ramp shape we consider the simple one-parameter families $r_1(x)=x+a\sin(2\pi x)$, and $r_2(x)=x+a\sin(4\pi x)$ with $\tmax = 1.25$ and $U=3$,
and compare the optimal values for the parameter $a$ 
obtained from the perturbative analysis to the Monte Carlo results.
The oscillation frequency of $r_1(x)$ is not compatible with the ramp time, while the oscillation frequency of $r_2(x)$ is identical to that found in the constrained optimized infinitesimal ramp (top left panel of Fig.~\ref{figconstrained}) and thus results in lower excitation energies. 

In Fig.~\ref{figoneparameter} the blue line with squares plots the excitation energy obtained from Eq.~(\ref{integral-E}) as a function of the shape parameter $a$. The red line with circles shows QMC results for the total energy which are shifted in such a way that the minima of the curves coincide. 
The necessary shifts $E_0=-0.279$ (for $r_1(x)$), and $E_0=-0.277$ (for $r_2(x)$) are  
still comparable to the ground state energy $E_0^{DMFT}=-0.284(3)$ obtained from a $T\rightarrow 0$ extrapolation of equilibrium DMFT energies for the $U=3$ model,
but the perturbative formula appears to underestimate the excitation energy of the system by $\delta E_0\approx 0.005-0.007$. Still, the variation of the excitation
energy as a function of the ramp-parameter $a$ is quite accurately reproduced. In particular, the perturbative analysis yields the correct values for the optimal parameters ($a_\text{opt}=0.175$ for $r_1(x)$ and $a_\text{opt}=0.87$ for $r_2(x)$), and thus the correct optimal ramp shape.   

For interactions $U\gtrsim 3.5$, the analytical prediction for the excitation energy is no longer in quantitative agreement with the Monte Carlo results. In the lower panel of Fig.~\ref{figoneparameter} we show Monte Carlo data for $U=4$ and $5$, with an arbitrary offset (chosen in such a way that the minima are at around the same energy). We see that as the interaction is increased, the minimum in the excitation energy shifts to smaller values of $a$, while the curvature at the minimum increases. This implies that the minimization of Eq.~(\ref{Eex}) yields ramp shapes which do not minimize the excitation energy, and which in the large-$U$ limit may be qualitatively wrong.

\subsection{Relaxation after the ramp}

We would finally like to address the relaxation dynamics after the ramp. If the goal of an experiment is to prepare an equilibrium state at given interaction strength, 
then it is not only the ramp time, but also the thermalization time, which determines the relevant time scale of this process. The relaxation dynamics after an (sudden) interaction quench has been studied in Refs.~\onlinecite{Moeckel2008, Eckstein09quench, Eckstein10quench}. These calculations demonstrated the existence of a ``dynamical phase transition" at some critical interaction strength $U_c^\text{quench}\approx 3.2$, which separates two qualitatively different relaxation regimes. After a quench to $U_c^\text{quench}$, the system thermalizes within 
a few inverse hoppings, whereas away from this particular interaction value thermalization occurs on much longer time scales.  

The critical interaction can be identified for example by plotting the  time evolution of the quantity\cite{Eckstein09quench}
\begin{equation}
\Delta n(t)=n(\epsilon_k=0_-,t)-n(\epsilon_k=0_+,t),
\end{equation}
which is the size of the discontinuity of the momentum distribution function $n(\epsilon_k,t)=\langle c^\dagger_{k,\sigma}(t)c_{k,\sigma}(t)\rangle$ at the Fermi energy. After a quench to $U<U_c^\text{quench}$, the system initially settles into a nonthermal quasistationary state characterized by a nonzero $\Delta n(t)$.\cite{Moeckel2008} For $U>U_c^\text{quench}$, $\Delta n(t)$ exhibits collapse-and-revival oscillations. At $U\approx U_c^\text{quench}$, $\Delta n(t)$ rapidly vanishes, with no sign of trapping in an intermediate nonthermal state.

\begin{figure}[t]
\centering
\includegraphics[width=1\linewidth]{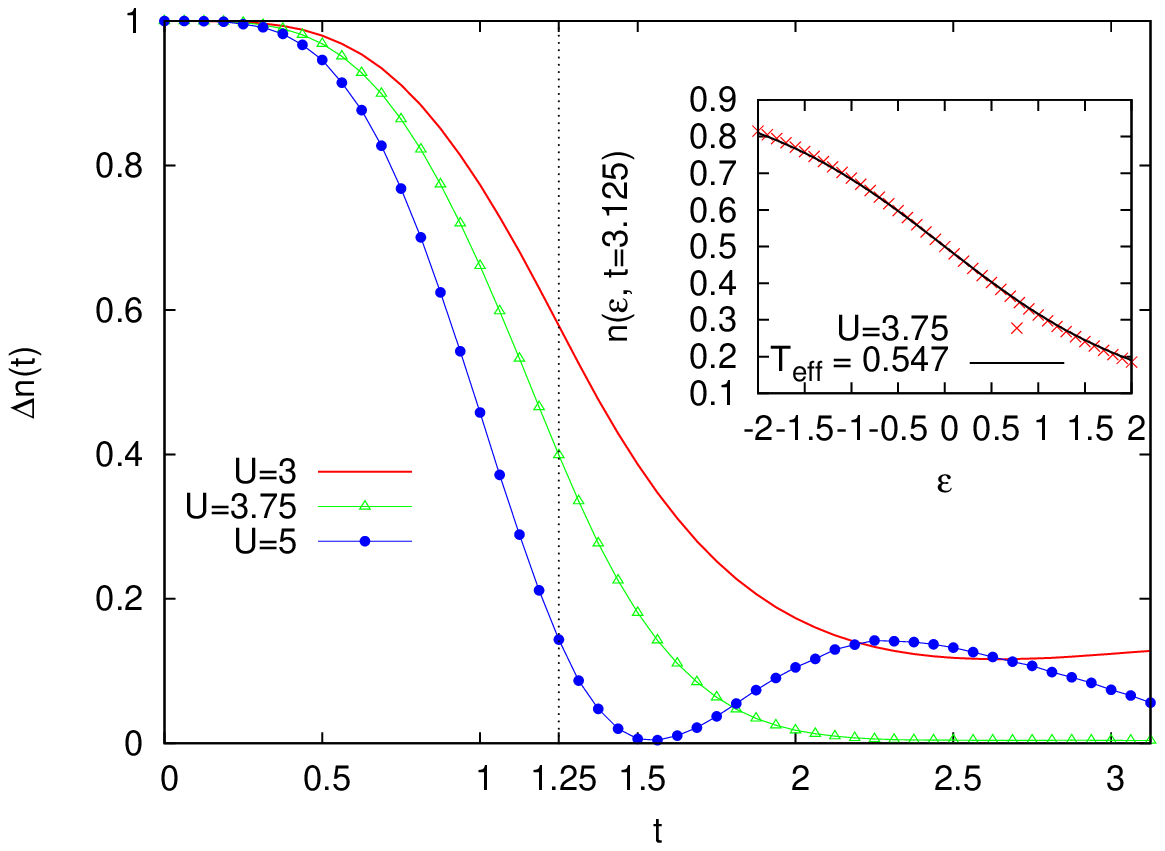}
\includegraphics[width=1\linewidth]{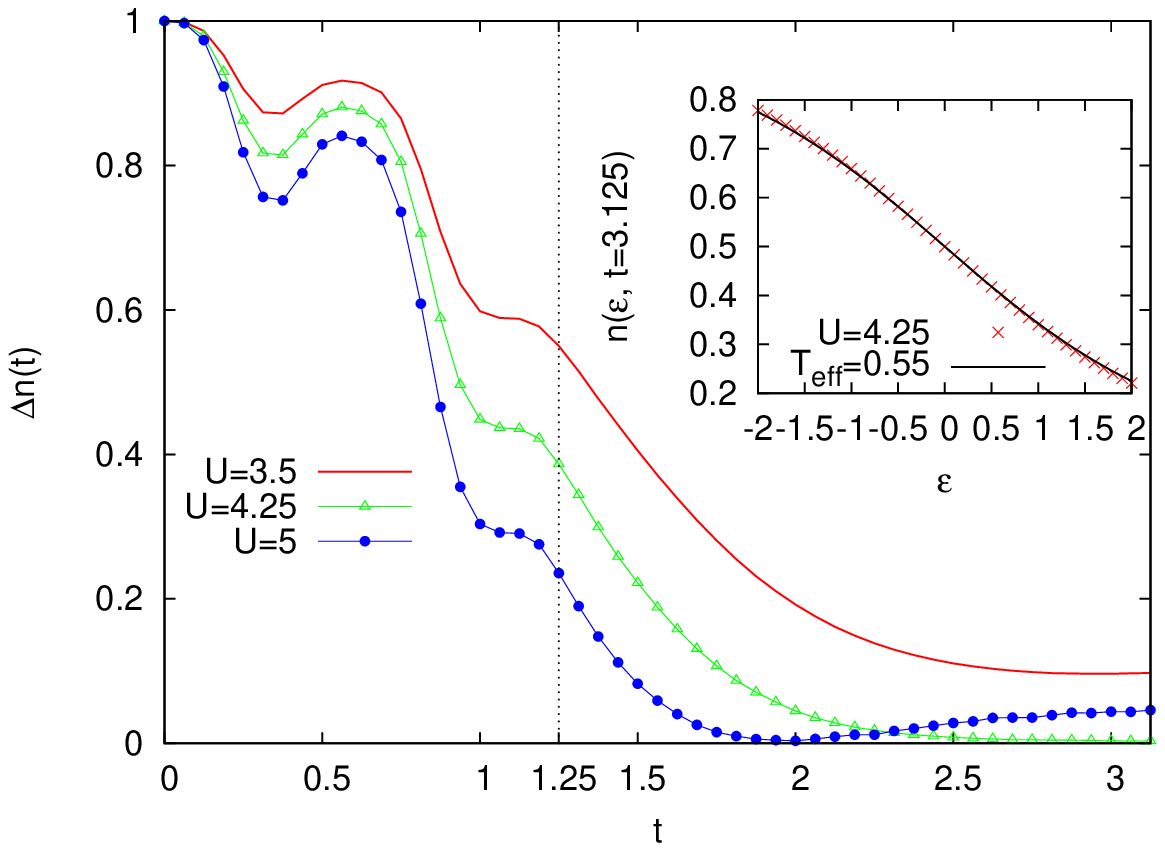}
\caption{Time evolution of the jump $\Delta n(\epsilon,t)$ in the momentum distribution function for indicated values of the ramp amplitude (ramp time $\tau=1.25$). The top panel shows results for linear ramps, and the bottom panel results for an oscillating ramp $r(x)=x+0.87\sin(4\pi x)$. For $U\approx 3.75$ (linear ramp) and $U\approx 4.25$ (optimized ramp), the jump vanishes within a time of less than $1.5$ inverse hoppings. The insets compare the distribution functions at $n(\epsilon, t=3.125)$ to the distribution functions of equilibrium systems with interaction $U$ and a total energy identical to the energy after the ramp (top panel: $U=3.75$, $T=0.547$, bottom panel: $U=4.25$, $T=0.55$).}
\label{figdelta}
\end{figure}

A very similar relaxation dynamics can be observed after an interaction ramp with relatively short ramp time $\tau$. In Fig.~\ref{figdelta} we plot $\Delta n(t)$ for different values of the interaction. The top panel shows results for linear ramps with ramp time $\tau=1.25$, and the lower panel for ramps of the form $r(x)=x+0.87\sin(4\pi x)$ with ramp time $\tau=1.25$. The critical interaction strengths (just before the onset of the collapse-and-revival oscillations) are $U_c^\text{linear}\approx 3.75$ and $U_c^\text{oscillating}\approx 4.25$. Apparently, the critical interaction strength for given ramp time $\tau$ depends on the ramp shape and this degree of freedom can be used to design protocols for which the desired interaction strength corresponds to a dynamical phase transition point. 

The green curves with triangles and the insets of Fig.~\ref{figdelta} demonstrate that also in the ramp case, the critical interaction strength is associated with fast thermalization. At the critical point, $\Delta n(t)$ vanishes within 
a few inverse hoppings, both for linear and oscillating ramps. In the insets we show a comparison of $n(\epsilon,t=3.125)$ with an equilibrium distribution function for the interacting system (temperature $T=0.547$ for the linear ramp and $T=0.551$ for the oscillating ramp). The temperatures of the thermalized systems have been computed by comparing the total energy after the ramp to the temperature-dependent total energy obtained from equilibrium DMFT simulations. The fast thermalization at the 
dynamical transition
and the ramp shape dependence of the critical interaction strength imply that it is possible to prepare thermal equilibrium states over a range of interaction values by suitably designing the ramp protocol. 

\section{Conclusion}
\label{outlooksection}
Using nonequilibrium DMFT, we investigated the excitation energy of the one-band Hubbard model after ramping up the interaction within a time $\tau$
of the the order of a few inverse hoppings. Based on the perturbative analysis of Ref.~\onlinecite{Eckstein10ramp} we determined optimal ramp 
protocols within a general set of piecewise linear ramp functions. This analysis indicates that the optimal ramp shape is oscillating around the linear ramp   
$U(t)/U(\tmax)=t/\tmax$, with a period which is determined by the support of the excitation density $R(\omega)$. (For small values of $U$, the latter 
depends mainly on the density of states.) Applying these optimized ramp shapes to larger interactions ($U=3$ for a model with bandwidth $4$) yields 
considerable reductions in the effective temperature, compared to linear ramps. For short ramp times $\tmax\lesssim 1.5$, both unconstrained paths 
and paths constrained to the interaction range $[0,U]$ result in effective temperatures which are 30\% to 50\% lower than those obtained with 
linear ramps. 

Up to intermediate interaction values, the optimal ramps determined from the perturbative approach coincide well with 
those determined from the QMC calculations. Our analysis thus suggest that one may use the perturbative formula to determine optimal ramps 
when an optimization using QMC simulations is not possible. This is true in particular for slow ramps with ramp times up to 100 inverse hoppings, which are used for the preparation of states in cold atom systems.  Guided by the optimal 
ramp shape at short times (oscillations superimposed on a linear ramp), one can, e.g., try to improve on the linear 
ramp by optimizing the amplitude $a$ in a ramp of the form $U(t)/U(\tmax)=t/\tmax + a \sin(\omega t)$. Using the 
``optimal'' frequency $\omega$ from Fig.~\ref{figoneparameter}b, for example, the perturbative approach predicts that 
an optimization of  $a$ results in a decrease of the excitation energy of more than 30\% with respect to the linear ramp, even in the limit of large $\tmax$,  where the absolute value of the excitation energy becomes small ($\sim 1/\tau^2$).

The precise shape (oscillation frequency) of the optimal ramps found in this study is specific to the $U=0$ initial state. For ramps
within the insulating phase, the ramp spectrum will be determined by the interaction energy $U$ rather than the bandwidth, and the 
optimal ramp shapes will be different. However, in view of our results a perturbative analysis  should still provide a numerically
efficient way to compute ramp shapes which result in low excitation energies.

In addition to minimizing the excitation energy, we have briefly addressed the question of thermalization after 
the ramp. We have demonstrated that the thermalization after a ramp strongly depends on the final interaction 
and the ramping parameters. Our results indicate the existence of a dynamical transition (associated with rapid 
thermalization), similar to the behavior after an interaction quench.  The ramp shape dependence of this dynamical 
transition point may be exploited to design ramp protocols which yield a thermal equilibrium state within a time of 
a few inverse hoppings after the switch-on of the interaction.

\acknowledgements
We thank N. Tsuji, C. Kollath, and P. Barmettler for useful discussions. The CTQMC calculations were run on the Brutus cluster at ETH Zurich using a code based on ALPS.\cite{ALPS} We acknowledge support from the Swiss National Science Foundation (Grant PP002-118866).

\end{document}